\documentclass[prl,twocolumn,aps]{revtex4}

\usepackage{amsmath}
\usepackage{amssymb}
\usepackage{latexsym}
\usepackage{amsfonts}
\usepackage{epsfig}
\usepackage{psfrag}
\usepackage{graphicx}

\usepackage{color}

\definecolor{black}{rgb}{0,0,0}
\definecolor{blue}{rgb}{0,0,1}
\definecolor{green}{rgb}{0,1,0}
\definecolor{red}{rgb}{1,0,0}

\newcommand{\f}[1]{\mbox{\boldmath$#1$}}

\newcommand{\bea}{\begin{eqnarray}}
\newcommand{\ea}{\end{eqnarray}}
\newcommand{\eea}{\end{eqnarray}}
\newcommand{\ord}{\,{\cal O}}

\begin{document}

{\large\bf Comment on: 
``Hawking Radiation from Ultrashort Laser Pulse Filaments''}

\vspace{.2cm}

In a recent paper \cite{Belgiorno} Belgiorno {\em et al} claimed to have 
observed the analog of the Hawking effect created by light pulses in silica 
glass because of the detection of radiation in a frequency range in which 
what they called ``phase horizons'' existed. 
Unfortunately, while the observations are very interesting, the cause of
the radiation is not understood, and we feel it is not justified to call 
this a detection of the Hawking effect in an analog system. 

The Hawking effect is the observation of thermal radiation 
(with the temperature being determined by the geometry)
from a time independent system, in which the radiation is caused by the 
(quasi) exponential tearing apart of the waves in the vicinity of the 
horizon. 
(This applies to black hole horizons, for white holes it would be the
time-reversed process, i.e., squeezing of waves.)  
Since the part of the (torn apart) wave-packet beyond the horizon can 
have negative energy, the other part may acquire positive energy and 
constitutes the Hawking radiation.
Out of these five conditions, only the last one
(i.e., the negative energy beyond the ``phase horizon'' which is 
related to the Landau criterion, see below) applies to the above 
experiment.  
Thus, it is an important step towards the observation of Hawking 
radiation, but not more. 

Even in the frame of the pulse, the system is not time independent -- 
the pulse itself lasts only a brief time 
(shorter than the time scale set by the surface gravity, 
i.e., Hawking temperature, see below)
and has space-time dependent sub-structure (due to the difference between 
the phase velocity of the radiation creating the pulse and the speed of 
the pulse itself). 
In the frame of the pulse, the photons co-moving with the pulse satisfying 
Eq.~(1) in \cite{Belgiorno} approach zero frequency due to 
$\omega^{\rm pulse}_{\rm frame}\approx
\omega^{\rm lab}_{\rm frame}-\f{v}_{\rm pulse}\cdot\f{k}$, 
making the creation of particles by that time dependence easy from an 
energetic point of view.  
(Momentum conservation is another matter.)
In fact, Eq.~(1) in \cite{Belgiorno} is closely related to the Landau 
criterion for particle creation $\omega^{\rm pulse}_{\rm frame}=0$. 

Second, there is no exponential tearing (or compaction) by the 
``horizon''. 
Since the group velocity of the photons under consideration is always
smaller than the speed of the pulse \cite{Belgiorno}, it just passes 
through them (i.e., there is no ``group velocity horizon''). 
As an instructive example, consider the dispersion relation of a 
massive particle $\omega^2=c^2k^2+m^2c^4/\hbar^2$. 
As in the experiment \cite{Belgiorno}, the phase velocity $\omega/k>c$
is larger than the group velocity $d\omega/dk<c$ and a perturbation 
with $v>c$ would have phase horizons but no group horizons. 
In a suitable Lorentz frame, this corresponds to an instantaneous, 
time-dependent perturbation with no horizon.

Third, the condition 
$\omega^{\rm lab}_{\rm frame}-\f{v}_{\rm pulse}\cdot\f{k}=0$ 
applies only to waves (in the frequency range of interest) which are moving 
in the same direction as the pulse. 
Thus Hawking radiation would also occur in this direction only. 
However, the authors of \cite{Belgiorno} observed photons at 90 degrees 
to the propagation direction of the pulse. 
The unpolarized nature of the observed radiation seems to rule out 
scattering of co-moving radiation as a source. 

As a fourth an final point, we note that the interpretation of this 
emission as Hawking radiation yields Hawking temperatures which are 
far too low to explain the observations. 
Even if the spectrum was deformed by dispersion and no longer Planckian, 
the following estimates for the energy and number of emitted photons would 
still yield the correct orders of magnitude.
In the frame of the pulse, the Hawking temperature would be given by 
the gradient of the speed of light $c$ in the medium 
$T_{\rm Hawking}=\hbar|\partial c/\partial x|/(2\pi k_{\rm B})$
which yields $\hbar|\partial\delta n/\partial x|/(2\pi k_{\rm B}n^2)$, 
where $n=1/c$ is the refractive index and $\delta n$ its change 
due to the Kerr effect.
Since the non-linearity $\delta n$ is quite small $\delta n\approx10^{-3}$,
this temperature could only create the observed photons at around 800~nm if 
one assumed that $\delta n$ changes on a sub-nm length scale, which seems 
unrealistic given that the wavelength of the photons generating the pulse 
is around 1000~nm.
Note that the transformation to the lab frame increases the frequency 
(though not the number) of the photons emitted in forward direction, but  
this Doppler shift does not apply to any photons emitted to the side. 

In addition, the Hawking effect would correspond to thermal radiation 
where the number of particles created goes as 
$\sigma A T^3_{\rm Hawking}(\Delta\vartheta)^2$. 
Here $A\sim L^2$ is the area of the horizon and $\Delta\vartheta$ 
is the solid angle into which the radiation is created.
Both are very small: due to the ``phase horizon'' condition, one has  
$\Delta\vartheta=\ord(10^{-2}\,{\rm rad})$ and the core size $L$ of a 
Bessel pulse with $7^\circ$ is a few $\mu$m. 
Together with $T_{\rm Hawking}\propto\delta n\approx10^{-3}$, one obtains  
estimates for the number of created particles which are several orders of 
magnitude too small to explain the data.  
An even simpler perturbation theory estimate for the maximum number of 
photons emitted per unit time scales with $(\delta n)^2$, showing serious 
difficulties with the interpretation as a stationary quantum process. 
Nevertheless, we admire the experimental technique of \cite{Belgiorno}
and we think that such a set-up may well provide the first observation of 
spontaneous Hawking emission in an analogue system.

\vspace{.2cm}

Ralf Sch\"utzhold$^{1,*}$ and William G.~Unruh$^{2,3,+}$
\\
$^1$Fakult\"at f\"ur Physik, Universit\"at Duisburg-Essen,
\\
D-47048 Duisburg, Germany 
\\
$^2$Department of Physics and Astronomy, University of 
British Columbia, Vancouver B.C., V6T 1Z1 Canada
\\
$^3$Institute for Theoretical Physics, Utrecht University, 
NL-3584 CE Utrecht, The Netherlands
\\
$^*${\tt ralf.schuetzhold@uni-due.de}, $^+${\tt unruh@phas.ubc.ca}

%%%%%%%%%%%%%%%%%%%%%%%%%%%%%%%%%%%%%%%%%%%%%%%%%%%%%%%%%%%%%%%%%%%%%%%%%%%%%%%
%%%%%%%%%%%%%%%%%%%%%%%%%%%%%%%%%%%%%%%%%%%%%%%%%%%%%%%%%%%%%%%%%%%%%%%%%%%%%%%
%%%%%%%%%%%%%%%%%%%%%%%%%%%%%%%%%%%%%%%%%%%%%%%%%%%%%%%%%%%%%%%%%%%%%%%%%%%%%%%
%%%%%%%%%%%%%%%%%%%%%%%%%%%%%%%%%%%%%%%%%%%%%%%%%%%%%%%%%%%%%%%%%%%%%%%%%%%%%%%

\end{document}